# Cosmic ray ionization and Jesse effect behavior in electrode systems with nanostructures: towards novel irradiation detectors and gas sensors

Changhua Zhan[1], Yuanzhi Pan[2], Zi Wang[1], Yanfang Wang[1], Zhongyu Hou[1]

[1] National Key Laboratory of Micro/Nano Fabrication Technology, Key Laboratory for Thin Film and Microfabrication of Ministry of Education, Research Institute of Micro/Nano Science and Technology, Shanghai Jiao Tong University, Shanghai 200030, People's Republic of China

[2] Central Academe, Shanghai Electric Group Co., Ltd.

E-mail: zhyhou@sjtu.edu.cn

## ABSTRACT

Gaseous electronic characteristics due to cosmic ray ionization in the electrode systems with ZnO and carbon nanostructures have been examined in atmospheric $Ar/N_2$ and $O_2/N_2$ mixtures. The nanostructures have been configured on one side of two electrode plates parallel to each other with ~280μm spacing. The results show that the discharge current increases with the applied voltage enduring three stages: linear stage, saturation stage, and sparking stage, same as the controlled samples without nanostructures except that the sparking criterion is significantly lower as expected. The other important phenomenon, rarely documented in the literatures, lies in that the electric conductivity at the quasi-linear stage is measured at the level of $10^{-11} \sim 10^{-10}$ S/m, 4~5 orders of magnitudes higher than that of the samples without nanostructures, which could be quantitatively construed by the $10^{8} \sim 10^{10}$ times increment of the cosmic ray ionization frequency. The results are interpreted based on the hypothesis that the role of one-dimensional nanostructures in this specific gaseous electronic phenomenon is based on the intensive field gradient effect, rather than the field enhancement effect. The field gradient due to the electric field flux convergence can lead to the polarization and capturing of the gas molecules by the biased nanostructures, the resulted inelastic collision may generate an appreciable population of metastable particles, which can result in the high yield of cosmic ray ionization.

**Key words:** Gaseous electronics, Ionization chamber, Cosmic rays, One-dimensional nanostructures, MEMS

# 1. Introduction

The experiments reported herein discuss the gaseous electronics interacting with one-dimensional nanoelectrode array (ONA) in the linear stage, when the applied voltage ($V$) is lower than the criterion of a self-sustaining discharge ($V_c$). The proper context of this work may include the physics of the ionization chamber without charge multiplication, where the *I-V* characteristics due to the ionizing irradiation include: $1/I$ increases with $1/V$ or $1/V^2$ and saturate at $1/I_s$ [1, 2] with $I_s = e \cdot v_i\, N\, d$, where $e$ is for elementary charge constant, $v_i$ is for the ionization frequency per unit volume due to the irradiation, e.g., $v_i \approx 3.5 \text{cm}^{-3}\text{s}^{-1}$ in atmospheric air for cosmic rays at low altitude [3], $N$ is for the gas number, and $d$ is for the electrode separation. Thus, it may be concluded that the ONA can impact $I_s$ only when it can impact $v_i$. How? In the context of the gaseous electronics interacting with ONAs, which is among the hot topics in the nanotechnology [4-7], the mechanisms of the field emission [8], the field ionization [7], and the electron multiplication [4, 6] have been scrutinized. Those three mechanisms generally do not lead to the increase in $v_i$, and the electronic manifestation of $v_i$ may be overlapped by them. This is because that the resulted charge flux and charge separation processes in the gases are different from that of the photon ionization behavior of the cosmic rays; and they should lead to the gaseous electronic characteristics, the exponential relationship as to the *I-V* characteristics, different from that of the quasi-linear *I-V* relationship as in a gas chamber [1, 2]. Consequently, how the ONAs impact $v_i$ could be examined by the *I-V* characteristics; and whether there is any impact could be measured by whether the gaseous conductivity extrapolated from the *I-V* relationship in an electrode system with ONAs (ESON) varies from that of the case without.

In this letter, the atmospheric gaseous electronic characteristics of the ESON samples with ZnO nanorods and carbon nanotubes (CNTs) well below $V_c$ have been examined. The gas conductivity due to the cosmic ray ionization in the linear *I-V* stage is found to be 4~5 orders of magnitudes larger than that of the samples without the ONAs. The phenomenological pattern has been illustrated by the generation of the metastable population due to the interaction between the gas molecule and the ONAs [6]. The results show that the electrode system with proper arrangement of one-dimensional nanostructures could be used to enhance the sensitivity of the ionizing irradiation detectors. Besides, the quasi-linear *I-V* characteristics could be considered as a novel alternative mechanism in the miniaturized gas sensors.

**2. Experiments**

The electrode configuration of the ESON samples is shown in figure 1(a), and the incorporated ONAs include three groups: dielectric barrier samples of CNT films covered with ethocel as shown in figure 1(b); the dense and sparse ZnO nanorods' samples as shown in figure 1(c) and 1(d), respectively. The ZnO nanorods are 6~8um in height, 20~80nm in diameter (*r*), and have been prepared through the well-known hydrothermal method [9]. The reaction temperature was 90°C and the reactants are zinc nitrate and hexamethylenetetramine. Before the growth process, the ZnO nanocrystals with 3~10nm in diameter have been prepared on the wafers as the seed layer through spin-coating. The CNT film (23~25um in thickness, $r$=5~7nm) was prepared with screen-printing method [10]. The electrode spacing is ~280μm, and the area of the ONAs region ($S$) is ~2.8mm$^2$. The devices under tests have been probed by a station of AutPri® HM-IP-EC12 in a vacuum chamber and the gas concentration is controlled by the mass flow controllers. The currents were measured using Agilent 2911A. The sample preparation and measurement instrumentation have been detailed in the supplementary data. Every device in a sample group has been repeatedly tested with $V$<210V in different atmospheric gases, and finalized with two rounds of sparking tests. The ONAs were positively biased during all the tests. To determine the impact of the electric conductivity of the polyimide spacer to the measurements, the controlled samples without nanostructures have been measured in different gas environments. The conductivity of the atmospheric air was measured by a Gerdien apparatus ZYKX® GLY-3G.

**3. Results and discussion**

The conductance of the gas gap ($C_{gp}$) is determined by the conductance of the gas ($C_{gs}$) and that of the polyimide spacer ($C_s$), connected in parallel. $C_s$ have been deduced from the $I$-$V$ measurements of the controlled parallel-plate sample without nanostructures (PPS), where the conductivity of the gas ($G_{gs}$) is measured to be $3.35 \times 10^{-15}$ S/m with the critical mobility at $3.2 \times 10^{-15}$ S/m. In fact, $G_{gs}$ could be considered as a constant and the most recent measurement result is 3~8$\times 10^{-15}$ S/m [11]. Thus, it is shown that $C_s \approx C_{gp} \approx 3.12 \times 10^{-12}$ S, given that $C_{gs} \approx 3.35 \times 10^{-17}$ S, which is trivial to the measured conductance. Therefore, $C_{gp}$ of a PPS is not gas sensitive in our instrumentation because the conductance of the polyimide spacer is not gas sensitive according to our measurements, where $C_s \approx C_{gp}$ appears to be a constant in different gases; and because the conductance of the gas is much smaller than the conductance of the spacer as $G_{gs} \ll G_s$ (conductivity of the spacer), although the conductance of the gas is sensitive. The conductance mechanism through the polyimide spacer may refer to the hopping model within the scope inside the solid entity [12]. However, the conductance of the ESON samples are strongly gas sensitive and the characteristic behavior is described as follows.

The characteristic *I-V* curves of the ESON samples and the controlled samples have been shown in figure 2, where the linear fitting is used to describe the *I-V* relationship before the saturation. To increase the readability, the *I-V* curves of the ZnO dense array samples have been offset by +2nA and those of the controlled samples have been offset by -2nA. The linear-fitted *I-V* curves of the ESON samples with different configurations in different gases, available in the supplementary data, are not straightforward in a compact figure so that the linearity has been expressed in conductivity, shown in figure 3, where the effect of $G_s$ has been considered. Four phenomenological issues of the results will be our focus: 1) $G_{gs}$ of the ESON samples is as high as $10^{-11} \sim 10^{-10}$S/m level; 2) Both the admixture of Ar and $O_2$ in $N_2$ possess higher $G_{gs}$ than the pure $N_2$ does; 3) $G_{gs}$ decreases from the peak value when the concentration of 'impurity' content ($C_{in}$) increases beyond certain criterion point ($C_{cr}$) corresponding to the $G_{gs}$ peak value; 4) The behavior of $G_{gs}$-$C_{in}$ relationship depends on the density of the ZnO nanorods and significantly different when the nanostructures are covered with dielectric film. The possible illustrations would be discussed as follows.

*3.1. Elevated gas conductivity*

The equivalent conductivity of the gases in an ESON is about $10^4 \sim 10^5$ times higher than that of the PPSs, or a Gerdien apparatus' results in the context of geophysics for low altitude air [11, 13]. Given that the gradient of the electric field in the vicinity of the nanostructures is $1.85 \times 10^{16} \sim 3.27 \times 10^{23} \text{V/m}^2$, the distance within which the field intensity ($E_l$) larger than $1 \times 10^6 \text{V/m}$ due to the enhancement effect of ONAs in the tested ESON samples is limited to ~95nm, less than two mean free paths of the electrons, according to the calculation of the Laplacian equation. Thus, the hypothesis about that the general role of the 1D nanostructures is the field enhancement induced electron multiplication which leads to multiple nanoscale corona discharges cannot explain the observations. Besides, the electron multiplication should lead to nonlinear *I-V* characters [14], which is inconsistent with the observations. The field enhancement effect of the ONAs may also lead to electron field emission (FE) at lower applied voltages; however, the FE electrons should not be included in the charge sources herewith, or the *I-V* curves should be nonlinear according to the *Fowler-Nordheim* formulae [15]. For the similar reason [16], the postulation of field ionization at low field intensity [16, 17] is also not possible herein. Thus, the ionization source in our measurements is limited to the cosmic rays, whose distribution are basically constant in the lower altitudes [11, 13, 18, 19], and $G_{gs}$ should have been in the order of $10^{-15}$ S/m, instead of $10^{-10}$ S/m. How to explain this increment?

It is suggested that an appreciable metastable population could be generated in the 'convergence band' region [20] through the collision processes concerning the intensely polarized gases due to the field gradients. If it is true, due to the metastable population, the cosmic ray ionization yield that can be described by the photon ionization cross section [21] will increase [22, 23]. Thus, $v_i$ in the gap is increased to $v_{ip}$ in average, suppose that the metastable population is generated near the ONAs [20] and diffuse in the whole gap. In the following, we shall analyze how $v_{ip}$ impacts $G_{gs}$.

Based on the electrodynamics' description of the drift current and the charge continuity, we could formulate the gaseous electronics herein as follows:

$$\vec{J_{e,l}} = -en_{e,i}\vec{\mu_{e,l}} \cdot \vec{E_l} \qquad (1,2)$$

$$\frac{\partial n_{e,i}}{\partial t} + \nabla \cdot (\frac{\vec{J_{e,l}}}{\mp e}) = v_{ip}dSn_0 - \beta n_e n_i \qquad (3,4)$$

where $\vec{J_{e,l}}$, $n_{e,i}$, $\mu_{e,i}$, is the current density, number density, drift mobility for electron and positive ions, respectively; $\beta$ is the recombination coefficient; $n_o$ is the gas number density; and $\vec{E_l}$ stands for the Laplacian field intensity. In the static condition and based on the polarity band postulation [6], the equations can be transformed into such a form:

$$\frac{dJ_e}{dz} = e\Gamma_i - \frac{\beta}{e \cdot \mu_i \cdot \mu_e \cdot E_l^2} J_e(J - J_e) \qquad (5)$$

where $J = J_e + J_i$ is the total current density, $\Gamma_i = v_{ip}d^2Sn_0$, and $z$ is the direction normal to the surface of the ONAs. In the context of ionization chamber [1, 2], the solution of linearity approximation could be obtained for $G_{gs}$:

$$G_{gs} \cong K_c \sqrt{\frac{4e^2\mu_i\mu_e v_{ip}dSn_0}{\beta}} \qquad (6)$$

where the boundary conditions have been $J_e=0$ and $J_i=J$ at the cathode and $J_i=0$ and $J_e=J$ at the anode, $K_c$ is the nonlinearity factor. According to equation 6, $G_{gs}$ will be larger if $v_{ip} > v_i$, which illustrate the elevated conductivity of the ESON samples. The quantitative predictions cannot be given herein because the detailed knowledge about the metastable population still needs further scrutiny. However, according to equation 6, $v_{ip}$ is ~$10^{10}$ times of $v_i$ by using the $G_{gs}$ measurement results. This implies that the increase of ionization frequency must be reasonably accounted for by the secondary processes, e.g., the photon-electrons via the Compton process [21] or the energetic photon emission via excitation processes, rather than the direct ionization [22, 23].

*3.2. Conductivity increment due to the admixtures*

It is expected that the ionization yield of cosmic rays in the mixture of different gases is larger than that of a pure gas, due to the *ad hoc* Penning ionization, referenced as Jesse effect sometimes [24-26]. As shown in figure 3(a) to 3(d), the conductivity of the dense and sparse ZnO nanorods' samples increases with the Ar and $O_2$ concentration in $N_2$ before 100~200ppm; this behavior is accord with the expectation of the Jesse effect [25]. The quantitative treatment could be deduced from the kinetic schemes of the related reactions, e.g., the rate coefficients [27] of the associative reactions between $N(^2P)$ or $N(^2D)$ with oxygen atoms are at the level of $10^{-12}cm^{-3}s^{-1}$. However, as shown in figure 3(a) and 3(c), $C_{cr}$ in the $O_2/N_2$ mixture in the samples of ZnO nanorods' sparse array is significantly different from that of the dense array. This implies that the Penning process herewith is sensitive to the property of the ONAs, i.e., sensitive to the field convergence characteristics [6]. Thus, the majority of the metastable population must be resulted from the processes due to the nanoelectrodes, instead of the cosmic ray impact. This also accounts for the larger increment due to the trace impurities comparing to that of a conventional Ar chamber without nanostructures [25].

*3.3. Conductivity reduction due to the admixtures*

It is suggested that there are two current limiting processes competitive to the charge flux increment due to the Penning ionization. The first one is the formation of the clusters of low charge number and high inertia through associative Penning ionization and charge transfer collisions between the clusters and the metastable atoms. In $Ar/N_2$ mixtures, the clusters can be held by Van de Waals force [28]; while in $O_2/N_2$, new molecules can be formed [27] and charge attachment of electrons may further modify the behavior of $C_{in}$-$G_{gs}$ relationship beyond $C_{cr}$. The second one is the collisional relaxation of the metastable particles, which will result in that the metastable population number density tends to decrease with the increase of $C_{in}$, i.e., $v_{ip}$ tends to decrease with $C_{in}$. As a result, those two processes may lead to the competitive effects to decrease charge flux so that the increase of $C_{in}$ beyond $C_{cr}$ could result in the decrease of $G_{gs}$ as shown in figure 3(a) to 3(d). The detailed differences between $O_2/N_2$ and $Ar/N_2$ in the same sample can be illustrated by the differences in the cluster chemistry.

*3.4. Conductivity in dielectric barrier samples*

Another interesting result is the behavior of the dielectric barrier samples of CNTs, shown in figure 3(e) and 3(f), where the concentrated charge density at the surface of the ethocel film covered on CNTs may lead to the process of Malter effect that can in turn increase the electron flux density. Thus, in this case, the charge clusters' formation is not only the competitive process that decrease the general conductivity, it is also favorable to the formation of the charge concentration because of their low mobility, and accordingly to intensify the secondary electron emission due to the Malter effect. This has resulted in that the $C_{cr}$ becomes larger and resulted in a behavior of $G_{gs}$-$C_{in}$ relationship contrary to those of the ZnO samples when $C_{in}$<900ppm. In the future works, we shall examine this phenomenon using ZnO nanorods' samples covered with $Al_2O_3$ dielectric thin films to exclude the differences in the materials being used.

Sparking tests have been performed to study how the cosmic ray ionization impact the breakdown of an ESON. First, the *I-V* curves of the ESON samples are shown to behave as switching devices as a PPS, although $V_s$ of an ESON sample is always lower. The sparks initiate when criterion ($V_s$) is reached where *I* increases abruptly to the protection limit (1μA) and noise and light are obvious. There is no transition stages ever observed as that of a corona discharge, although the field non-uniformity, measured by *d/r*, is about $10^4$, satisfies the condition for corona discharges. This is contrary to the expectations of the nanoscale corona discharge hypothesis. Second, as shown in figure 4(b), $V_s$ of the same sample deviates from that of the first sparking tests shown in figure 4(a), significantly. The deviations may be explained by the formation of new structures, or by the splashed fragments due to the strike of the sparks which may decrease the gap spacing. The results show that the sparking tends to cause irreversible changes to the device structures. Recall the discharge behavior of an ESON with *d*=6~12μm, where strong evidences of electron multiplication processes without sparking have been recorded [20], the results herein implies that longer gaps should cause the atmospheric gas discharges in an ESON tends to behave similarly to that of a common parallel-plate system. This could account for the formation of high current density (2.14μA/cm$^2$) stage of the sparse ZnO nanorods' sample after the first sparking test, where the sparks may lead to a limited region of smaller gap size.

**4. Summary and conclusion**

In a conclusion, the electric conductivity of pure and mixture gases can be significantly enhanced by one-dimensional nanostructures biased with voltages much lower than the criterion of self-sustained discharges. The tentative illustration of the observed phenomenological pattern based on the hypothesis of polarization induced processes show that a theory emphasizing the intense field gradient induced by the flux convergence effect of 1D nanoelectrodes' array may lead to more flexible and self-consistent models to give accounts for the diverse and correlated pattern of the observations. As to the possible application opportunities of this study, the detection of irradiation rays may take the advantages of the high ionization yield in an ESON to improve the detection limit by elevating the saturation current [1, 2], or miniaturize the device dimensions under the similar sensitivity. As to the future efforts on the theoretical scrutiny, we will focus on the possible relationship between the dimensions of an ESON sample, especially the dimension of the gas gap, with the uniqueness of this specific phenomenon. As a preliminary proof, our expanded experiments show that the increment of the gaseous conductivity in the quasi-linear stage tends to become trivial in an ESON with smaller gap spacing of several micrometers, or with larger gap spacing of several tens of centimeters.

**Acknowledgements**

This work was supported by the Natural Science Foundation (Grant Nos. 60906053 and 09ZR1415000) and Ministry Foundation (Grant Nos. 51308050309 and 9140A26010112JW0301).

**References**

[1] Greening J R 1964 *Phys. Med. Biol.* **9** 143-54
[2] Scott P B and Greening J R 1963 *Phys. Med. Biol.* **8** 51-7
[3] Millikan R A 1932 *Phys. Rev.* **39** 391
[4] Modi A, Koratkar N, Lass E, Wei B and Ajayan P M 2003 *Nature* **424** 171-4
[5] Sadeghian R B and Kahrizi M 2008 *IEEE Sensors* **8** 161-9
[6] Hou Z, Cai B and Liu H 2009 *Appl. Phys. Lett.* **94** 163506
[7] Sadeghian R B and Islam M S 2011 *Nature Mater.* **10** 135-40
[8] Zhang W, Fisher T and Garimella S 2004 *J. Appl. Phys.* **96** 6066-72
[9] Greene L E, Law M, Goldberger J, Kim F, Johnson J C, Zhang Y, Saykally R J and Yang P 2003 *Angew. Chem. Int. Edn* **42** 3031-4
[10] Hou Z, Cai B, Liu H and Xu D 2008 *Carbon* **46** 405-13
[11] Pawar S D, Murugavel P and Lal D M 2009 *J. Geophys. Res.* **114** D02205
[12] Kim T, Kim W, Lee T, Kim J and Suh K 2007 *Express Polym. Lett.* **1** 427-32
[13] Dhanorkar S and Kamra A K 1992 *J. Geophys. Res.* **97** 20345-60
[14] Yang J 1983 *Gas Discharges* (Beijing: Science Publication) p 24
[15] Fowler R H and Nordheim L 1928 *Proc. R. Soc.* **119** 173-81
[16] Liu X and Orloff J 2005 *J. Vac. Sci. Technol.* B **23** 2816
[17] Goodsell A, Ristroph T, Golovchenko J and Hau L V 2010 *Phys. Rev. Lett.* **104** 133002
[18] Locher G L 1932 *Phys. Rev.* **39** 883
[19] Carmichael H and Steljes J F 1955 *Phys. Rev.* **99** 1542
[20] Hou Z, Zhou W, Wang Y and Cai B 2011 *Appl. Phys. Lett.* **98** 063104
[21] Millikan R A and Anderson C D 1932 *Phys. Rev.* **40** 325
[22] Usoskin I G and Kovaltsov G A 2006 *J. Geophys. Res.* **111** D21206
[23] Cassé M and Goret P 1978 *Astrophys. J.* **221** 703-12
[24] Siska P E 1993 *Rev. Mod. Phys.* **65** 337-412
[25] Jesse W P and Sadauskis J 1952 *Phys. Rev.* **88** 417
[26] Gillen K T, Jones P R and Tsuboi T 1986 *Phys. Rev. Lett.* **56** 2610-3
[27] Kossyi I, Kostinsky A Y, Matveyev A and Silakov V 1992 *Plasma Sources Sci. Technol.* **1** 207
[28] Bouchiat C, Bouchiat M and Pottier L 1969 *Phys. Rev.* **181** 144

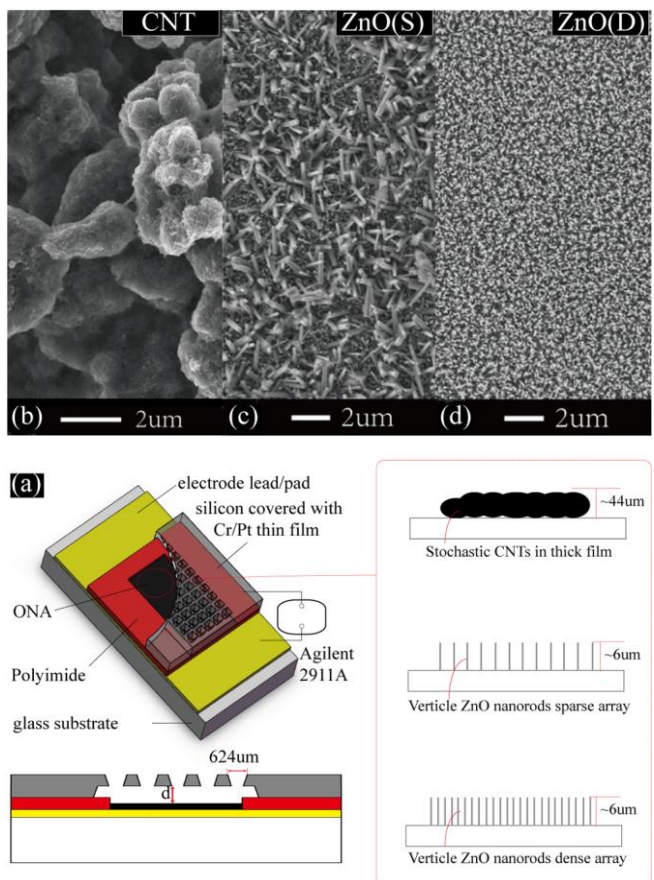

Figure 1. (Color online) (a) Electrode configuration schematics; Scanning electron microscopy images of (b) CNT film, ZnO nanorods' (c) sparse array, and (d) dense array.

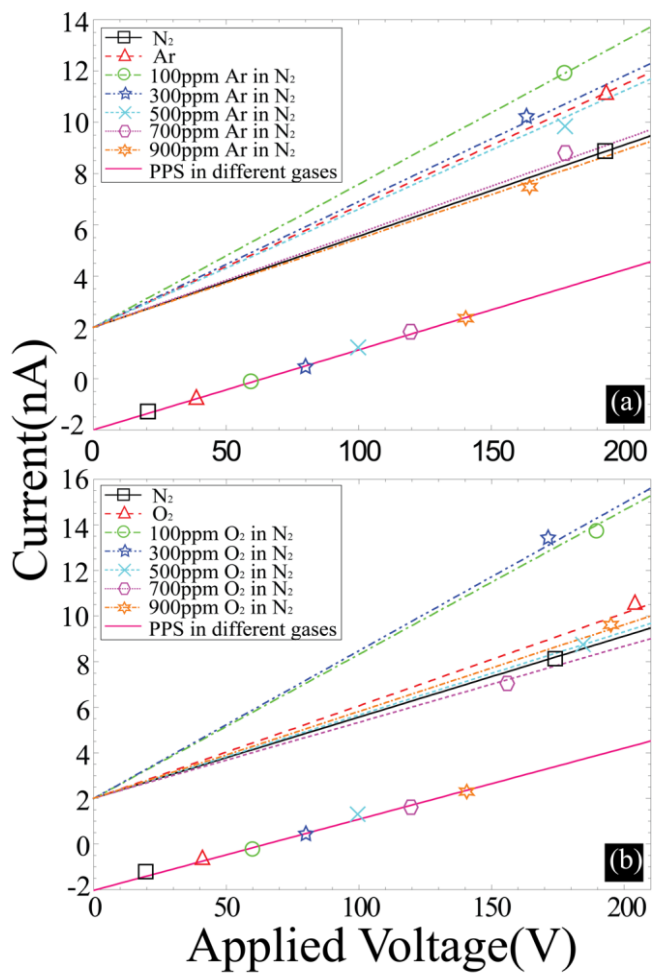

Figure 2. (Color online) *I-V* curves of the dense ZnO nanorods' samples and the controlled samples in (a) Ar/N$_2$ and (b) O$_2$/N$_2$.

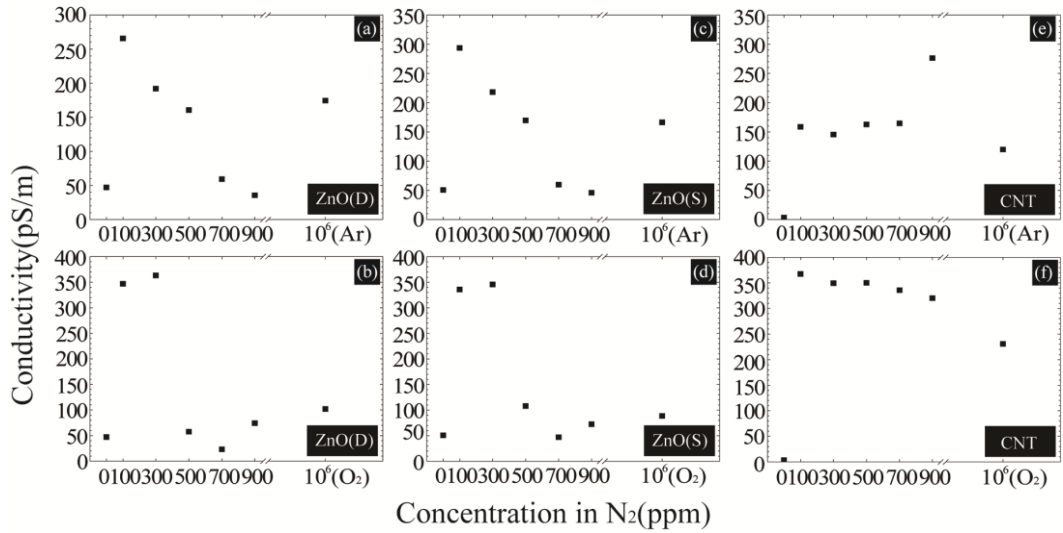

Figure 3. Gas conductivity in different gases.

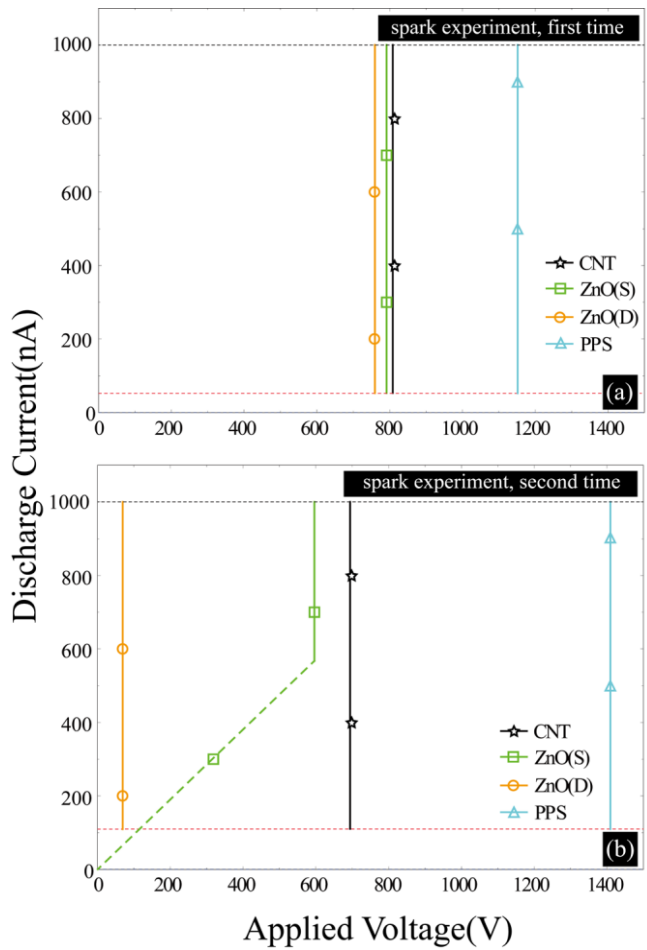

Figure 4. (Color online) *I-V* characteristics of spark tests.

16